\documentclass[journal=jacsat,manuscript=article]{achemso}
\usepackage{chemformula}
\usepackage[T1]{fontenc} 
\usepackage{amssymb,amsmath,graphics,graphicx,float,braket}
\usepackage{epstopdf}
\usepackage{array,multirow,setspace}
\usepackage[autostyle]{csquotes}
\usepackage{dcolumn}
\usepackage{bm}
\usepackage{physics}\usepackage{subscript}
\usepackage{soul}
\usepackage{subfigure}
\usepackage[utf8x]{inputenc} 
\usepackage{verbatim}
\usepackage{comment}

\title{Electrical and magneto transport in 2D semiconducting MXene \ch{Ti2CO2}}
\author{Anup Kumar Mandia}
\affiliation{Department of Electrical Engineering, Indian Institute of Technology Bombay, Powai, Mumbai-400076, India}
\author{Namitha Anna Koshi}
\affiliation{Indo-Korea Science and Technology Center (IKST), Jakkur, Bengaluru 560065, India}
\author{Bhaskaran Muralidharan}
\affiliation{Department of Electrical Engineering, Indian Institute of Technology Bombay, Powai, Mumbai-400076, India}
\author{Seung-Cheol Lee}
\affiliation{Electronic Materials Research Center, KIST, Seoul 136-791, South Korea}
\email{leesc@kist.re.kr}
\author{Satadeep Bhattacharjee}
\affiliation{Indo-Korea Science and Technology Center (IKST), Jakkur, Bengaluru 560065, India}
\email{s.bhattacharjee@ikst.res.in}


\begin{document}
\begin{tocentry}
\includegraphics[scale=0.45]{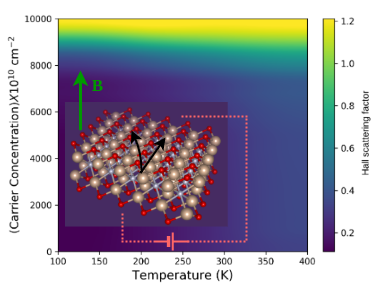}
\end{tocentry}
\begin{abstract}
The Hall scattering factor is formulated using Rode's iterative approach to solving the Boltzmann transport equation in such a way that it may be easily computed within the scope of ab-inito calculations. Using this method in conjunction with density functional theory based calculations, we demonstrate that the Hall scattering factor in electron-doped Ti$_2$CO$_2$ varies greatly with temperature and concentration, ranging from 0.2 to around 1.3 for weak magnetic fields. The electrical transport was modelled primarily using three scattering mechanisms: piezoelectric scattering, acoustic scattering, and polar optical phonons. Even though the mobility in this material is primarily limited by acoustic phonons, piezoelectric scattering also plays an important role which was not highlighted earlier. 
\end{abstract}

\newpage
\section{Introduction}
Since the advent of graphene \cite{novoselov2004electric}, two-dimensional (2D) materials have become the focus of intensive research due to their novel electrical, chemical, optical and mechanical properties \cite{mas20112d, akinwande2017review, novoselov2005two, neto2009electronic,R1}. Many other 2D materials like h-BN \cite{jin2009fabrication, zeng2010white, ci2010atomic}, transition metal dichalcogenides (TMDs) \cite{wilson1969transition, mattheiss1973band, wang2012electronics, duerloo2012intrinsic, geim2013van} and Xenes \cite{li2014black, molle2017buckled, carvalho2016phosphorene, mannix2018borophene, kong2017elemental, tao2017antimonene, zhang2018recent, tao2019emerging}, group V graphyne~\cite{R2} have been fabricated and deeply investigated for nanoelectronics application. 

Recently, MXene, a new class of 2D transition metal carbides, nitrides and carbonitrides have been synthesized \cite{barsoum2000mn+, barsoum2011elastic, naguib2011two, naguib2012two, naguib201425th, ghidiu2014synthesis} from $M_{n+1}AX_{n}$ phases (where \textquote{M} represents an early transition metal, \textquote{A} represents group IIIA or IVA element and \textquote{X} represents C and/or N, and n=1-3) by selective exfoliation of \textquote{A} atoms as the M-A bonds are much weaker than the M-X-M bonds. $M_{n+1}AX_{n}$ belong to the family of layered compounds with $P6_{3}/mmc$ symmetry. Other possible pathways of synthesizing MXenes are Due to the etching process during synthesis, MXenes are terminated with -O, -OH and -F groups. So, functionalized MXenes are generally written as $M_{n+1}X_{n}T_{x}$, with T standing for the surface terminating groups. Zha et al. \cite{zha2015role} have investigated the mechanical, structural and electronic properties of \ch{M2CT2} MXenes and shown that surface functional groups (T = F,OH,O) have considerable impact on the crystal structure of these materials. Oxygen functionalized MXene structure possess smaller lattice parameter and also shows higher mechanical stability as compared to fluorine and hydroxyl functionalized ones. They also have higher thermodynamic stability than fluorine and hydroxyl functionalized ones \cite{khazaei2014two, xie2014role, peng2014unique}.\\

The family of 2D materials has been quickly expanding since the discovery of the first MXene in 2011. Several MXenes have been synthesised to date, and their characteristics have been extensively studied by various researchers. Chemical stability, hydrophilic behaviour, strong electronic conductivity, and outstanding mechanical qualities are among the characteristics they exhibit. Numerous experimental and theoretical works have presented the potential applications of MXene in various fields like energy storage \cite{tang2012mxenes, naguib2012mxene, hu2014two, xie2014role, liang2015sulfur, er2014ti3c2, xie2013extraordinarily}, gas sensors \cite{chen2015co, yu2015monolayer}, bio-sensors \cite{liu2015novel}, adsorbents \cite{peng2014unique, mashtalir2013intercalation}, supercapacitor \cite{ghidiu2014conductive, lukatskaya2013cation}, water treatment \cite{peng2014unique}, biomedical application \cite{anasori20172d, liu2015novel}, electromagnetic interference shielding \cite{shahzad2016electromagnetic, han2016ti3c2} and so on. Some of these applications are related to high electronic conductivity and there is a need to explore it. From experimental and theoretical investigations, it is known that the metallic or semiconducting nature of MXenes depend on the surface termination group and a vast majority of the  members of this family are metallic \cite{anasori20172d, lukatskaya2013cation, wang2017first, azofra2016promising, djire2019electrocatalytic}. Also, $Ti_{3}C_{2}T_{x}$ is the most conductive one among all the MXenes synthesized. Metallic conductivity of MXenes can be engineered by controlling their surface chemistry or intercalation mechanism \cite{hart2019control}.
\par
In this work, we work we have performed an comprehensive study of the electronic and magneto-transport in 
Ti$_2$CO$_2$ using first principles  based transport calculations. A key quantity for studying the carrier concentration and drift mobility in a semiconductor is the so called Hall factor~\cite{nag1975galvanomagnetic,hl1}, which is commonly considered to be equal to one, which implies Hall mobility and the drift mobility are the same. But in reality, in many materials it differs from one which lead to wrong estimation of the carrier density and the drift mobility. In the present work, we formulate the Hall factor in a simple form which can be calculated within the framework of Rode's iterative scheme of solving the Boltzmann transport equation. We apply such a scheme to understand the temperature dependence of the Hall factor in Ti$_2$CO$_2$ with the inputs obtained from DFT based simulations. Furthermore, even though the transport properties of Ti$_2$CO$_2$ have been addressed in the previous studies, the role of piezoelectric scattering was never discussed. We also demonstrate that  piezoelectric scattering plays an important role in this material.
It should be mentioned here that we have excluded the effect of spin-orbit coupling for the moment. For MXenes containing heavy 4d and 5d transition metals, the relativistic spin-orbit coupling (SOC) affects the electronic structures significantly~\cite{khazaei2014two}. There are few known exceptions to it in MXene literature. One such example is, the BB phase of Ti2CF2, which demonstrate multiple Dirac cones and giant spin orbit coupling. The effect of  SOC on the electronic structure of pristine/bare Ti2C is investigated by B. Akgenc \textit{et al.}~\cite{ACK}. For the minimum energy 1T phase
of Ti2C, the effect of  SOC is minute whereas it becomes significant for 2H-Ti2C. It modifies dispersion of the bands arising from d-orbitals. We have considered the 1T phase for modeling Ti2CO2 and hence we expect the effect of SOC to be small.

\subsection{Structural model}
We start with the investigation of structural properties of \ch{Ti2C}. Keeping up with transition metal dichalcogenides (TMD) notation, two phases 1T and 2H of MXenes are considered \cite{structures}. Both 1T and 2H phases have hexagonal symmetry with C atom sandwiched between two Ti triangular lattices. In the 1T phase, the transition metal atoms are not in line (side view - Figure 1(a)). 
The symmetry group of 1T-\ch{Ti2C} is $P-3m_{1}$ (No. 164) \cite{akgencc2020phase, wang2022dft}.
For 2H structure, transition metal atoms are in line (side view - Figure 1(c)) or they are stacked on top of each other 
and has the symmetry group $P-6m_{2}$ (No. 187). 
From the total energy calculations, 1T phase is lowest in energy for non-magnetic (NM) configuration. To obtain the correct magnetic ground state, we calculate the energies of the different magnetic orderings (ferromagnetic-FM, antiferromagnetic-AFM1 and antiferromagnetic-AFM2). For AFM calculations, we construct 2$\times$1 supercell consisting of four transition metal atoms and the corresponding AFM configurations are given in SI. In AFM1, intralayer coupling is ferromagnetic and interlayer configuration is antiferromagnetic ordering. In the case of AFM2, the intralayer coupling is antiferromagnetic and interlayer ordering is ferromagnetic. From comparison of total energies, AFM1 ordering is preferred for 1T-\ch{Ti2C} whereas the lowest energy configuration is FM for 2H-\ch{Ti2C}. This is consistent with reported literature \cite{}. The magnetic moment of FM configuration of 2H-\ch{Ti2C} is 2 $\mu_{B}$/cell. The optimized lattice constants of 1T- and 2H-\ch{Ti2C} are 3.06 and 3.05 \AA{} respectively. The corresponding thickness of the layer are 2.30 and 2.47 \AA{} respectively and match with previous DFT reports \cite{akgencc2020phase}. The 1T phase of \ch{Ti2C} is semiconducting in nature whereas the 2H phase is half-metallic (details are given in SI).
Further, we study the structure of oxygen functionalized 1T-\ch{Ti2C} and the most stable configuration is given in Figure 1. Here, we see that the oxygen atom on top lies in line with Ti atom in the lower layer and vice versa. This structure is non-magnetic and belongs to the spacegroup P-3m1 (No. 164). The optimized lattice parameter is 3.03 \AA{} which agrees with other literature \cite{zha2016thermal}. The thickness of the layer (calculated as the distance between the two oxygen atoms) is around 4.45 \AA{}.
\section{\textit{ab-initio} calculations}
Electronic structure calculations are carried out using density functional theory (DFT) implemented in plane wave code, Vienna ab-initio Simulation Package (VASP). For pseudopotentials, the projector augmented wave (PAW) approach is used. The exchange-correlation functional is treated using generalized gradient approximation (GGA) parameterized by Perdew-Burke-Ernzerhof (PBE) formalism. The plane wave cut off energy is set to 500 eV. The conjugate gradient algorithm is used for structural optimization. The convergence criteria for energy and force are $10^{-6}$ eV and -0.01 eV/\AA{} respectively. A vacuum of thickness 20 \AA{} along the z-direction is employed to avoid interactions between the neighboring layers and a Monkhorst-Pack k-mesh of 17$\times$17$\times$1 is used for Brillouin zone sampling. The DFT-D2 method is used for van der Waals correction. The crystal structures are visualized using VESTA \cite{momma}. Phonon spectrum is calculated using VASP in combination with Phonopy software. Here, we employ a supercell of size 4$\times$4$\times$1 and a 3$\times$3$\times$1 k-mesh to determine the dynamical matrix.
\subsection{Electronic structure and lattice dynamics}
The electronic band structure of \ch{Ti2CO2} with the corresponding density of states (DOS) is given in Figure 2. It is semiconducting with a band gap of 0.25 eV with the valence band maximum (VBM) and conduction band minimum (CBM) at $\Gamma$ and M respectively.To determine the dynamical stability, phonon dispersion spectrum of \ch{Ti2CO2} is calculated and presented in Figure 2. There are no imaginary frequencies in the phonon spectrum hence \ch{Ti2CO2} can be experimentally realized as free standing layer. \\

\subsection{\textit{ab-initio} parameters needed for the transport calculation}
The calculated \textit{ab-initio} parameters obtained using DFT calculations are reported in the Table-\ref{table1}. The acoustic deformation potential as well as elastic moduli are calculated along the out-of-plane (ZA), longitudinal (LA) and transverse (TA) directions using the method described in Zha \textit{et al}~\cite{zha2016thermal}. The piezoelectric and dielectric constants (both high and low frequency) are calculated using density functional perturbation theory (DFPT)~\cite{DFPT}. 

\section{Methodology: Solution of Boltzmann Transport Equation using Rode's iterative method}
Transport coefficient calculation are performed by using our tool AMMCR \cite{mandia2021ammcr, mandia2019ab}. Brief methodology of solving the Boltzmann Transport Equation (BTE) is presented below.
\subsection{Case I: Carriers are in electric field}
BTE for the electron distribution function f is given by
\begin{equation}
\frac{\partial \textit{f(\textbf{k})}}{\partial t} + \textbf{v}\cdot\nabla _rf + \frac{e\textbf{E}}{\hbar} \cdot\nabla _kf= \frac{\partial \textit{f}}{\partial t}\Bigr|_{\substack{coll}},
\label{BTE}
\end{equation}

where $e$ is the electronic charge, \textbf{v} is the carrier velocity, \textbf{E} is the applied electric field, \textit{f} describes the probability distribution function of carrier in real and momentum space as a function of time, $\frac{\partial \textit{f}}{\partial t}\Bigr|_{\substack{coll}}$ represents the change in the distribution function with time due to collisions. Under steady state , $\frac{\partial \textit{f(\textbf{k})}}{\partial t} = 0$ and spatial homogeneous condition ($\nabla _rf=0$), equation \ref{BTE} can be written as 

\begin{equation}
\frac{e\textbf{E}}{\hbar}\cdot\nabla _kf = \int [ s(k,k')f(1-f') -  s(k',k)\; f'(1-f)] dk' ,
\label{BTE1}
\end{equation}

where $s(k,k')$ represents the transition rate of an electron from a state $k$ to a state $k'$.  At lower electric fields, the distribution function is given by \cite{rode1970electron, rode1970electron1, rode1975low}

\begin{equation} 
f(k) = f_0[\epsilon(k)] +  g(k)cos\theta,
\label{dist_ft}
\end{equation}

where $f_0[\epsilon(k)]$ is the equilibrium distribution function, \(g(k)\) is perturbation in the distribution function, and \(\cos\theta\) is the angle between applied electric field and \(k\). Higher order terms are neglected here, since we are calculating mobility under low electric field conditions. Now perturbation in the distribution function \(g(k)\) is required for calculating the low-field transport properties. The perturbation in the distribution function \(g(k)\) is given by \cite{rode1970electron, rode1970electron1, rode1975low} 

\begin{equation}
g_{k,i+1} = \frac{S_i(g_k,i)- v(k)(\frac{\partial f}{\partial z})- \frac{eE}{\hbar}(\frac{\partial f}{\partial k}) } {S_o(k)} .
\label{pert}
\end{equation} 

where $S_{i}$ represents in-scattering rates  due to the inelastic processes  and $S_{o}$ represents the sum of out-scattering rates. $S_o=\frac{1}{\tau_{in}(k)}+\frac{1}{\tau_{el}(k)}$. Where $\frac{1}{\tau_{el}(k)}$ is the sum of the momentum relaxation rates of all elastic scattering processes and  $\frac{1}{\tau_{in}(k)}$ is the momentum relaxation rate due to the in-elastic processes.

The expression for  $\tau_{el}(k)$, $S_{i}$ and $\frac{1}{\tau_{in}(k)}$ are given by the following equations
\begin{equation}
\frac{1}{\tau_{el}(k)} = \int (1 - X) s_{el}(k,k') dk'
\label{el}
\end{equation}

\begin{equation}
\ S_i(g_k,i) = \int X g_{k',i} [s_{in}(k',k)(1-f) + s_{in}(k,k')f ]dk'
\label{in_sc}
\end{equation}

\begin{equation}
\frac{1}{\tau_{in}(k)}=\int [s_{in}(k,k')(1 - f') + s_{in}(k',k)f']dk'
\label{out_sc}
\end{equation}

where X is the cosine of the angle between the initial and the final wave vectors, $s_{in}(k,k')$ and $s_{el}(k,k')$ represents transition rate of an electron from state $k$ to $k'$ due to inelastic and elastic scattering mechanisms respectively. Since, $S_{i}$ is function of g(k), thus equation \ref{pert} is to be calculated iteratively \cite{rode1975low}. In our previous work we have used the same procedure to calculate mobility for ZnSe \cite{mandia2019ab} and CdS\cite{mandia2021ammcr}. Drift mobility $\mu$ is then calculated by the following expression \cite{mandia2019ab} 

\begin{equation}
\mu = \frac{1}{2E} \frac{\int v(\epsilon) D_s(\epsilon) g(\epsilon) d\epsilon}  
{\int D_s(\epsilon) f( \epsilon )  d\epsilon },
\label{mobility}
\end{equation}
\quad

where $D_S(\epsilon)$ represents density of states. The carrier velocity is then calculated directly from the ab-initio band structure by using the following expression

\begin{equation}
\ v(k) = \frac{1}{\hbar} \frac{\partial{\epsilon}}{\partial{k}}.
\label{velocity}
\end{equation}

From these, we can evaluate the electrical conductivity given as

\begin{equation}
\ \sigma = \frac{n e \mu_e}{t_z},
\label{sigma}
\end{equation}
where $n$ is the electron carrier concentration, $t_z$ is the thickness of the Ti$_2$CO$_2$ layers along z-direction which is 4.45\AA~ in this case.
\subsection{Case II: Carriers are in both electric and magnetic field}
A similar method, also introduced by Rode \cite{rode1973theory} to solve BTE under arbitrary magnetic field. In this case distribution function is given by \cite{rode1973theory}

\begin{equation} 
f(k) = f_0[\epsilon(k)] +  x g(k) + y h(k),
\label{dist_ft_bf}
\end{equation}

where h(k) represents perturbation in distribution function due to magnetic field, and y is direction cosine from $ \textbf{B} \times \textbf{E} $ to $\textbf{k}$, where $B$ is applied magnetic field. Substituting equation \ref{dist_ft_bf} in equation \ref{BTE}, we get a pair of coupled equations that can be solved iteratively \cite{rode1973theory} 
  
\begin{equation}
g_{i+1}(k) = \frac{S_i(g_{i}(k))- \frac{eE}{\hbar}(\frac{\partial f}{\partial k}) + \beta S_i(h_{i}(k))} {S_o(k)(1 + \beta^2) } .
\label{pert_g}
\end{equation} 
 
\begin{equation}
h_{i+1}(k) = \frac{S_i(h_{i}(k))+ \beta \frac{eE}{\hbar}(\frac{\partial f}{\partial k}) - \beta S_i(g_{i}(k))} {S_o(k)(1 + \beta^2) } .
\label{pert_h}
\end{equation}

where $\beta = \frac{e v(k) B}{\hbar k S_o(k)}$.
The above expression shows that the perturbations to the distribution function due to the electric field ($g$) and magnetic field ($h$) are coupled to each other through the factor $\beta$ and the in scattering rates $S_i$. It should be highlighted that such a representation cannot be obtained using standard relaxation time approximation (RTA) and can only be seen using the current method.
The components of conductivity tensor in terms of perturbations are given by  
\begin{equation}
\sigma_{xx} =   \frac{e \int v(\epsilon) D_s(\epsilon) g(\epsilon) d\epsilon}{2E} 
\label{sigma_xx}
\end{equation}

\begin{equation}
\sigma_{xy} =   \frac{e \int v(\epsilon) D_s(\epsilon) h(\epsilon) d\epsilon}{2E} 
\label{sigma_xy}
\end{equation}

The Hall coefficient $R_{H}$, Hall mobility $\mu_H$ and Hall factor $r$ are respectively calculated by 
\begin{equation}
R_{H} =  \frac{\sigma_{xy}}{B(\sigma_{xx}\sigma_{yy} + \sigma_{xy}^2)} 
\label{hall_coeff}
\end{equation}

\begin{equation}
\mu_{H} =  \sigma_{xx}(0)|R_{H}|
\label{hall_mobility}
\end{equation}

\begin{equation}
r =  \frac{\mu_{H}}{\mu}
\label{hall_factor}
\end{equation}

where $\sigma_{xx}(0)$ is value of $\sigma_{xx}$ in the absence of the magnetic field.

\subsection{Scattering Mechanisms}
\subsubsection{Acoustic Scattering}
The scattering rates due to the acoustic phonons can be expressed as~\cite{zha2016thermal}

\begin{equation}
\frac{1}{\tau^j_{ac}(E)} = \frac{D^2_{jA} k_{B} T k}{\hbar^2 C^j_{A} v} 
\label{acoustic_rate}
\end{equation}

where $T$ is temperature, $ k $ is the wave vector, $ C_{A} $ is the elastic modulus, $ \hbar $ is the reduced Planck's constant, $D_{jA}$ is acoustic deformation potential for the $j^{th}$ acoustic mode and $ k_{B} $ is Boltzmann constant. $v$ is the group velocity of the electrons. $j\in$ LA,TA,ZA. The energy dependence here enters through the wave-vector k. We have
performed an analytical fitting of the lowest conduction band with a six-degree polynomial to get smooth curve for group velocity which help us to get a one by one mapping between the wavevector k and the band energies.

\subsubsection{Piezoelectric Scattering}

The piezoelectric Scattering rates~\cite{kaasbjerg2013acoustic} are calculated as follows,

\begin{equation}
\frac{1}{\tau_{pz}(E)} = \frac{1}{\tau_{ac}(E)} \times \frac{1}{2}\times \left( \frac{e_{11} e}{\epsilon_0 D_{A}}\right) ^2 
\label{acoustic_rate}
\end{equation}

where $e_{11}$ is piezoelectric constant (unit of C/m), $\epsilon_0$ is vacuum permeability.   
\subsubsection{Polar Optical Phonon (POP) Scattering}
The polar optical phonons are the source of inelastic scattering in the system. The inelastic scattering rates from a given $\bf k$-state are given in terms of $S_{in}$ (in) and $S_o$ scattering as in the Eq. \ref{in_sc} and Eq. \ref{out_sc}. 

The \textit{out scattering contribution} due to polar optical phonon scattering is given by \cite{nag2012electron, kawamura1992phonon}
\begin{equation}
\begin{aligned}
\frac{1}{\tau_{in}(k)}= & \frac{C_{pop}}{(1- f_{0}(E))} [N_V (1 - f_{0}(E + \hbar \omega_{pop}) I^+(E) \frac{k^+}{v(E + \hbar \omega_{pop})}\\  
            &+ (N_V+1) (1 - f_{0}(E - \hbar \omega_{pop}) I^-(E)\frac{k^-}{v(E - \hbar \omega_{pop})}]
\end{aligned}
\label{out_sc_pop}
\end{equation}

where 
\begin{equation}
I^{+}(E) = \int_0^{2\pi} \frac{1}{q_{a}} d\theta 
\label{J_plus}
\end{equation}

\begin{equation}
I^{-}(E) = \int_0^{2\pi} \frac{1}{q_{e}} d\theta 
\label{J_plus}
\end{equation}

\begin{equation}
q_{a} = \left( k^2 + \left( k^{+}\right)^2 - 2k k^{+} cos \theta \right)
\label{q_ab}
\end{equation}

\begin{equation}
q_{e} = \left( k^2 + \left( k^{-}\right)^2 - 2k k^{-} cos \theta \right)
\label{q_em}
\end{equation}

where $\theta $ is angle between initial wave vector $k$ and final wave vector $k^{'}$, $k^+$ and $k^-$ represents wave vector at energy $E+ \hbar \omega$ and $E- \hbar \omega$ respectively.

\begin{equation}
C_{pop} = \frac{e^2 \omega_{pop}}{8 \pi \hbar \epsilon_0} \times \left( \frac{1}{\kappa_{\infty}} - \frac{1}{\kappa_0} \right) 
\label{pop_const}
\end{equation}

$\kappa_{\infty}$ and $\kappa_{0}$ represents high frequency and low frequency dielectric constant.
 
The \textit{in scattering contribution} due to polar optical phonon scattering can be represented by the sum of in-scattering due to absorption and emission of polar optical phonons 

\begin{equation}
S_i^{in}(k) = S_a^{in}(k) + S_e^{in}(k)
\label{in_sc_pop}
\end{equation}

where $S_a^{in}(k)$ represents in-scattering due to absorption of polar optical phonon from energy $E-\hbar \omega_{pop}$ to energy $E$ and $S_e^{in}(k)$ represents in-scattering due to emission of polar optical phonon from energy $E+\hbar \omega_{pop}$ to energy $E$.
   
\begin{equation}
S_{a}^{in}(k) = C_{pop} (N_V+1) f_{0}(E)  J^-(E) \frac{k^-}{v(E- \hbar \omega_{pop}) f_{0}(E - \hbar \omega_{pop})}
\label{ab_in_sc_pop}
\end{equation}

\begin{equation}
S_{e}^{in}(k) = C_{pop} (N_V) f_{0}(E)  J^+(E) \frac{k^+}{v(E+ \hbar \omega_{pop}) f_{0}(E + \hbar \omega_{pop})}
\label{ab_in_sc_pop}
\end{equation}

\begin{equation}
N_V = \frac{1}{exp(\hbar\omega_{pop}/k_B T) - 1}
\label{N}
\end{equation}

\begin{equation}
J^{+}(E) = \int_0^{2\pi} \frac{cos \theta}{q_{i,e}}  d\theta 
\label{J_plus}
\end{equation}

\begin{equation}
J^{-}(E) = \int_0^{2\pi} \frac{cos \theta}{q_{i,a}} d\theta 
\label{J_plus}
\end{equation}

\begin{equation}
q_{i,a} = \left( \left( k^{-}\right)^2 + k^2 - 2k k^{-} cos \theta \right)
\label{q_ab}
\end{equation}

\begin{equation}
q_{i,e} = \left( \left( k^{+}\right)^2 + k^2  - 2k k^{+} cos \theta \right)
\label{q_em}
\end{equation}

While driving the expression for different scattering rate we have replaced term by $\frac{\hbar k}{m^*}$ by group velocity \cite{mandia2019ab, mandia2021ammcr}. The group velocity will be calculated directly from the DFT band structure \cite{mandia2019ab, mandia2021ammcr}.   

\begin{table} 
\caption{ Material Parameters used for $Ti_2CO_2$}
\begin{tabular}{cccc}
Parameters  &    Values  \\
\hline
PZ constant, $e_{11} $ (C/m)      &   $3 \times 10^{-13}$ \\
Acoustic deformation potentials, $D_{A} (ev):$       &    \\
$D_{A,LA} $  & 8.6     \\
$D_{A,TA} $  & 3.5     \\
$D_{A,ZA} $  & 0.7     \\
Elastic modulus, $C_{A} (N/m) $ & 		\\
$C_{A,LA}  $    & $301.7  $      \\
$C_{A,TA}  $    & $391.6 $      \\
$C_{A,ZA}  $    & $59.3 $      \\
Polar optical phonon frequency $\omega_{pop} (THz)$ &  \\
$\omega_{pop,LO} $  & 3.89     \\
$\omega_{pop,TO} $  & 3.89     \\
$\omega_{pop,HP} $  & 8.51     \\
High frequency dielectric constant, $ \kappa_{\infty}$     &   23.570798  \\
Low frequency dielectric constant, $ \kappa_{0}$     &    23.6 \\

\end{tabular}
\label{table1}
\end{table}

\section{Results and Discussion}
Let us discuss about electron transport first, before we go on to magnetotransport. Also, in order to fully understand magnetotransport in this material, these results must be comprehended.
\\
\subsection{Electronic transport}
In the Fig.\ref{scattering_rate}(a), we show the scattering rates due to the phonons (acoustic and optical) and due the piezoelectric scattering. It can be seen that the most dominant contribution is due the acoustic phonons, followed by the piezoelectric scattering. The POP scattering has the least effect. Therefore the conductivity or mobility in Ti$_2$CO$_2$ is limited by acoustic phonons. To understand the nature of the acoustic phonons that limits the conductivity in Ti$_2$CO$_2$, we compute the scattering rates due to individual acoustic phonons which are shown in the Fig.\ref{scattering_rate}(b). From this figure, it can be understood that LA phonons are the ones which play key role here. Therefore the two main carrier scattering mechanisms involved here are due to LA phonons and piezoelectric scattering.
To realize the trend of mobility with respect to temperature and electron concentration, let us consider Matthiessen's rule
\begin{equation}
\frac{1}{\mu}=\frac{1}{\mu_{ac}}+\frac{1}{\mu_{pz}}+\frac{1}{\mu_{pop}}
\label{Matt}
\end{equation}
where $\mu$ denotes total mobility and the suffixes ac, pz, and pop denote the acoustic, piezoelectric, and polar optical contributions, respectively. In the Fig.\ref{mobility} (a) we show the temperature variation of mobility at different electron concentration. As expected the mobility decrease with increasing electron concentration as well as temperature. The mobility values are not very different for electron concentration within the range [$1\times 10^{11}$ cm$^{-2}$ - $1\times 10^{13}$ cm$^{-2}$] however there is a drastic decrease of the mobility which can be seen from the Fig.\ref{mobility}(b) at electron concentration of $1\times 10^{14}$ cm$^{-2}$ where the mobility drops rapidly. We fit the temperature dependent mobility was  through a power law model as follows,
\begin{equation}
\mu=AT^{-\gamma}
\label{power}
\end{equation}
here A is the prefactor and $\gamma$ is the power law exponent. The exponent depends on the concentration. With increase in carrier concentration from $10^{11} cm^{-2}$ to $10^{14} cm^{-2}$ the exponent changes from 1.4 to 0.7. We fit the mobility data to the Eq.\ref{power} within the temperature region 100-700K. The values of the exponents look similar to values seen in standard electron doped semiconductors particularly when we consider the scattering mechanisms are dominated by acoustic phonons~\cite{STO,Vining}. This explains also the nature of the longitudinal conductivity shown in the Fig-\ref{conductivity}.  The conductivity here evolve as a result of  competition between two terms: (1) The increase in conductivity due to the increase in carrier concentration and (2) the decrease in conductivity (due to the decrease in mobility) with increase in carrier concentration. The latter has a temperature dependence ($\mu=AT^{-\gamma}$). As we have shown above the exponent $\gamma$ has small value (about 0.7) for the carrier concentration of about $10^{14} cm^{-2}$, there is a crossing of the conductivity curves that can be seen in the Fig-\ref{conductivity} (a) for the concentration $10^{13} cm^{-2}$ and $10^{14} cm^{-2}$.

In the Fig.\ref{mo-co}, we show the contribution of different components of mobility and their temperature evolution. In the left panel ( Fig.\ref{mo-co}(a)) we show both piezoelectric as well as phonon contribution (acoustic and optical) to the mobility. According to the Eq.\ref{Matt} above, the reciprocal nature of the relationship between total mobility and individual components results in the case where the component with the least value is the most significant. From the figure, it is understood that LA acoustic mode is the dominant contribution at all the temperature. However, there is a significant contribution of piezoelectric scattering something that was not addressed by other studies on the same material.
\subsection{Magnetotransport: The Hall factor}
In many situations the Hall factor is assumed to be equal to one, then we have $\mu_H=\mu$. This is true when $R_H=\frac{1}{ne}$. The Hall and the drift mobility are same in this case. This is the case when one assumes constant relaxation time and the band is parabolic. However, for a general case, the Hall factor can therefore be expressed as 
\begin{equation}
r=ne R_H=\frac{ne}{B}\frac{\sigma_{xy}}{(\sigma_{xx}\sigma_{yy}+\sigma_{xy}^2)}
\sim \frac{ne}{B}\frac{\sigma_{xy}}{\sigma_{xx}^2}
\label{factor}
\end{equation}
The above expression results from the fact that $\sigma_{xy}$ is very small and its square can be neglected in the denominator. Also we consider $\sigma_{xx}\sim\sigma_{yy}$. Combining Eq.\ref{sigma_xx}, Eq.\ref{sigma_xy} and Eq. \ref{factor}, the Hall factor can be expressed as,
\begin{equation}
r=\dfrac{n}{B}2E\frac{\int v(\epsilon)D_s(\varepsilon)h(\varepsilon)d\varepsilon}{[\int v(\epsilon)D_s(\varepsilon)g(\varepsilon)d\varepsilon]^2}
\label{r}
\end{equation}
The above formulation shows  the Hall factor is directly proportional to the carrier concentration, and the strength of the electric field, while it is inversely proportional to the magnetic field. In our formulation temperature dependence   enters via the perturbations to the distribution function $g(\varepsilon)$ and $h(\varepsilon)$ which in tern depend on the temperature dependent scattering rates. Recently, Macheda \textit{et al.} have calculated the Hall scattering factor in graphene~\cite{Factor1}. They also formulated Hall factor in terms of the solutions of the Boltzmann transport equation. However, they used RTA and the effect of magnetic field on the distribution function was not considered in an explicit way as we have done. In our case we use two perturbations $g(\varepsilon)$ and $h(\varepsilon)$ which are coupled to each other as can be seen in the equations \ref{pert_g} and \ref{pert_h}.  It can also seen from those equations that coupling between $g(\varepsilon)$ and $h(\varepsilon)$ further enhances at larger magnetic field through the factor $\beta$. For even small inelastic scattering, the coupling between $g(\varepsilon)$ and $h(\varepsilon)$ could be considerably large due to effect of magnetic field. In the scenario of larger magnetic fields, we expect our technique to perform better than the one proposed by Macheda \textit{et al.}~\cite{Factor1}.

It should be noted here that, if we drop the inelastic contribution which means that $S_i$=0 and $S_o=\frac{1}{\tau_{el}}$, in this case we stop at the zeroth iteration (i=0, in Eq.\ref{pert_g} and Eq.\ref{pert_h}). The situation is equivalent of doing calculations within RTA, (equivalent to the work done by Macheda \textit{et al}). It is vital to note, however, that unlike compound semiconductors, POP scattering should be almost absent in the case of a free-standing graphene, making the estimate of Macheda \textit{et al.} quite justifiable.

In the Fig.\ref{Hall-cond} (a) we show the Hall-conductivity with respect to the temperature. The value of the magnetic field used is  $0.4~\text{T}$ along the z-direction. The figure shows highest Hall conductivity at the electron concentration of $10^{13}\text{cm}^{-2}$

.In the Fig.\ref{hall_factor}(a) we show the variation of the Hall factor with the temperature, while in the Fig.\ref{hall_factor}(b) we show how Hall factor changes with doping concentration. It can be seen that the temperature dependence of the Hall factor is higher for the high doping concentration. The Hall factor varies within the range [0.1-0.5] up to concentration $10^{11}-10^{13} \text{cm}^{-2}$. The Hall factor rises steeply within the concentration range $10^{13}-10^{14} \text{cm}^{-2}$. At concentration, $10^{14}$ Hall factor reaches value 1.3 which is comparable to that of bulk Si. The same behaviour can be understood from the heat-map of the Hall scattering factor at various temperatures and concentrations shown in the Fig.\ref{Phase} where the results are shown for two values of the magnetic field 0.4T and 0.8T. It can be seen that the Hall scattering factor is as small as 0.2 for small doping and temperature. Therefore one has be cautions about estimating the carrier concentration and drift mobility in those regions. Use of Hall factor as one may lead to the overestimation of the carrier concentration and underestimation of the drift velocity.

\section{Conclusions}
In conclusion, we have studied the electrical and magneto-transport properties in electron doped semiconducting MXene Ti$_2$CO$_2$ using a combined approach of Rode's iterative scheme with DFT based methods. The electronic and vibrational properties needed as inputs are obtained from DFT simulations. Hall factor shows large deviation from unity as function of temperature and carrier concentration. At low doping and temperature the Hall factor is as low as 0.2 while at higher temperature and carrier concentration, the Hall factor is bigger than one which also depend on the value of magnetic field considered. This suggest that one has to take precautions while measuring the carrier concentration and drift mobility in such systems.

\clearpage
\newpage
\bibliography{Reference}
\subsubsection{}


\section{Acknowledements}
 AKM and BM gratefully acknowledge funding from Indo-Korea Science and Technology Center (IKST), Bangalore.

\section{Conflict of interest}
The authors declare that they have no conflict of interest.
\section{Author contributions}
SB conceived the idea. SB and NAK performed the DFT calculations. The transport calculations using AMMCR code were performed by AKM under supervision of SB. The manuscript was written by SB, NAK and AKM. BM and SCL contributed in discussions.
\clearpage

\begin{figure}
\subfigure[$ $]{\hspace{-0.5cm}\includegraphics[scale=0.5]{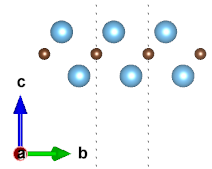}} 
\subfigure[$ $]{\hspace{1.5cm}\includegraphics[scale=0.5]{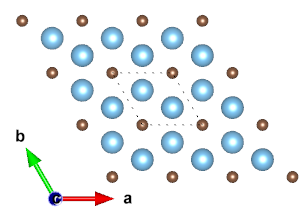}} \\
\subfigure[$ $]{\hspace{-0.25cm}\includegraphics[scale=0.5]{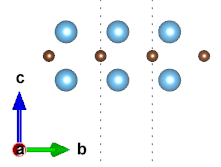}}
\subfigure[$ $]{\hspace{1.5cm}\includegraphics[scale=0.5]{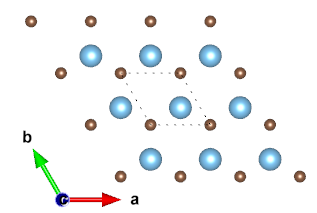}} \\
\subfigure[$ $]{\hspace{0.40cm}\includegraphics[scale=0.55]{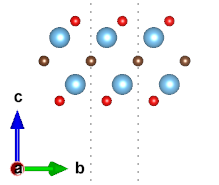}}
\subfigure[$ $]{\hspace{1.5cm}\includegraphics[scale=0.55]{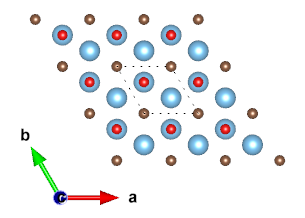}} \\
 \caption{Side view and top view of (a-b) 1T and (c-d) 2H phases of \ch{Ti2C} and (e-f) \ch{Ti2CO2}. Blue, brown and red balls correspond to Ti, C and O respectively.}
 \label{figure 1}
\end{figure}
\clearpage

\begin{figure}
\subfigure[$ $]{\hspace{-1cm}\includegraphics[scale=0.45]{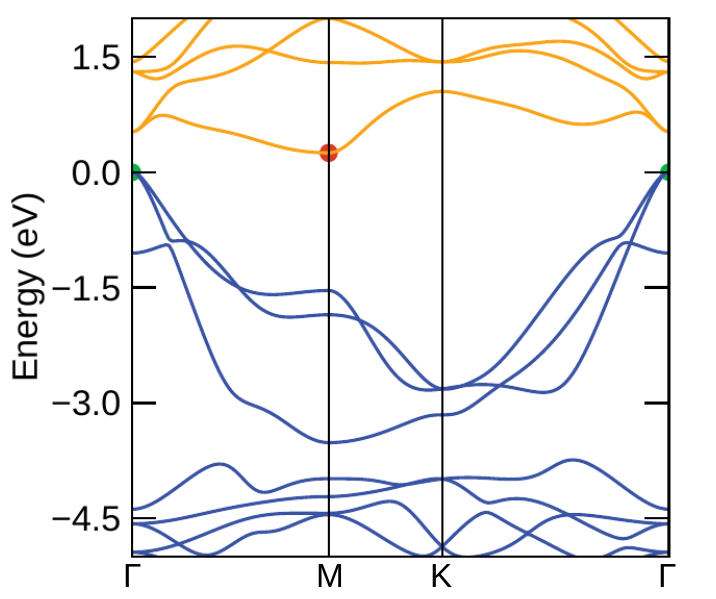}} 
\subfigure[$ $]{\hspace{1.2cm}\includegraphics[scale=0.45]{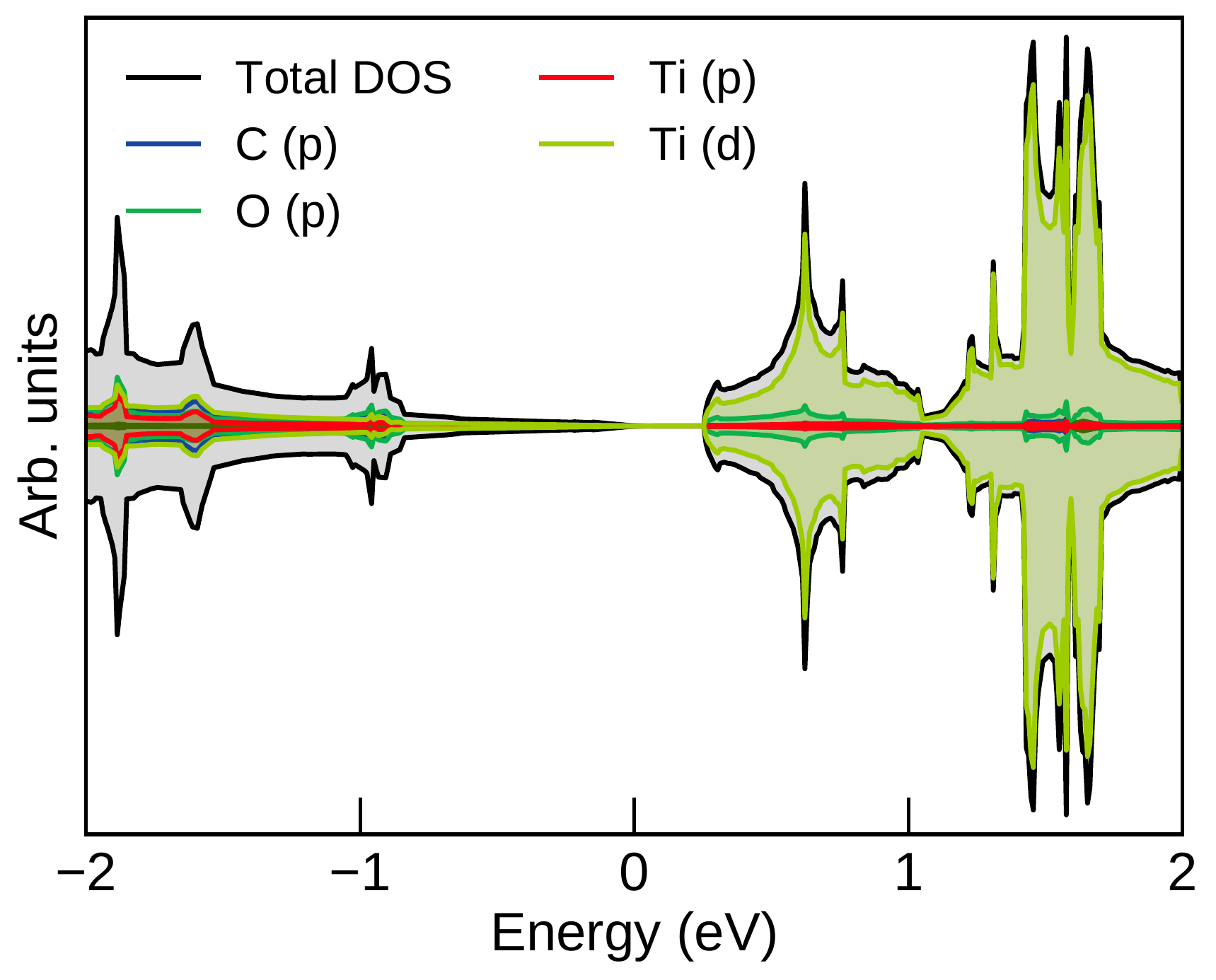}}\\
\subfigure[$ $]{\hspace{-0.5cm}\includegraphics[scale=0.5]{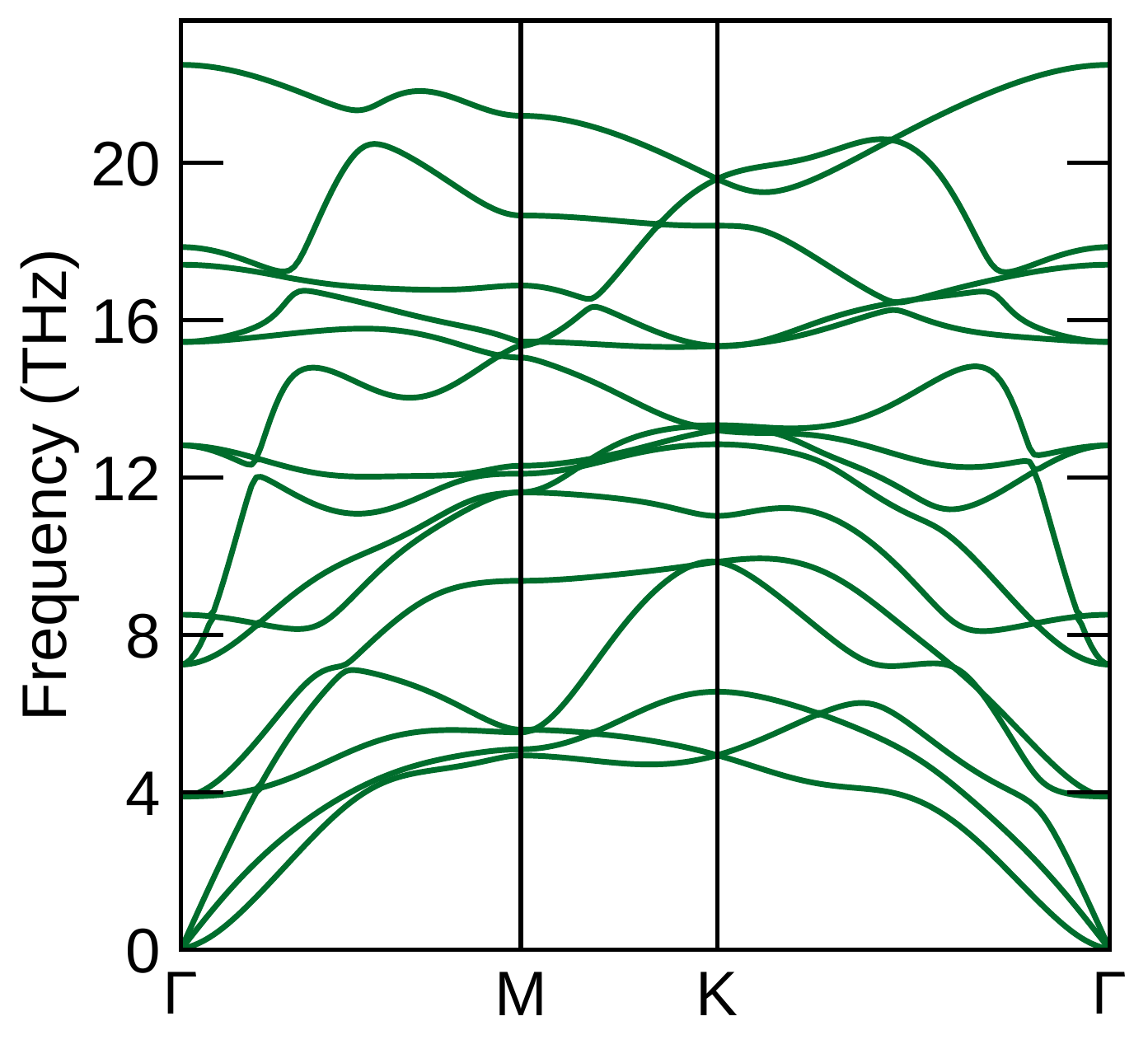}} 
 \caption{(a) Band structure of \ch{Ti2CO2} with the VBM and CBM denoted by green and red points respectively, (b) projected density of states and (c)phonon dispersion of \ch{Ti2CO2} along the high symmetry directions of the Brillouin zone.}
 \label{figure 2}
\end{figure}
\begin{figure}[hbt!]
    \centering
    \subfigure[$ $]{\includegraphics[width=80mm,height=60mm]{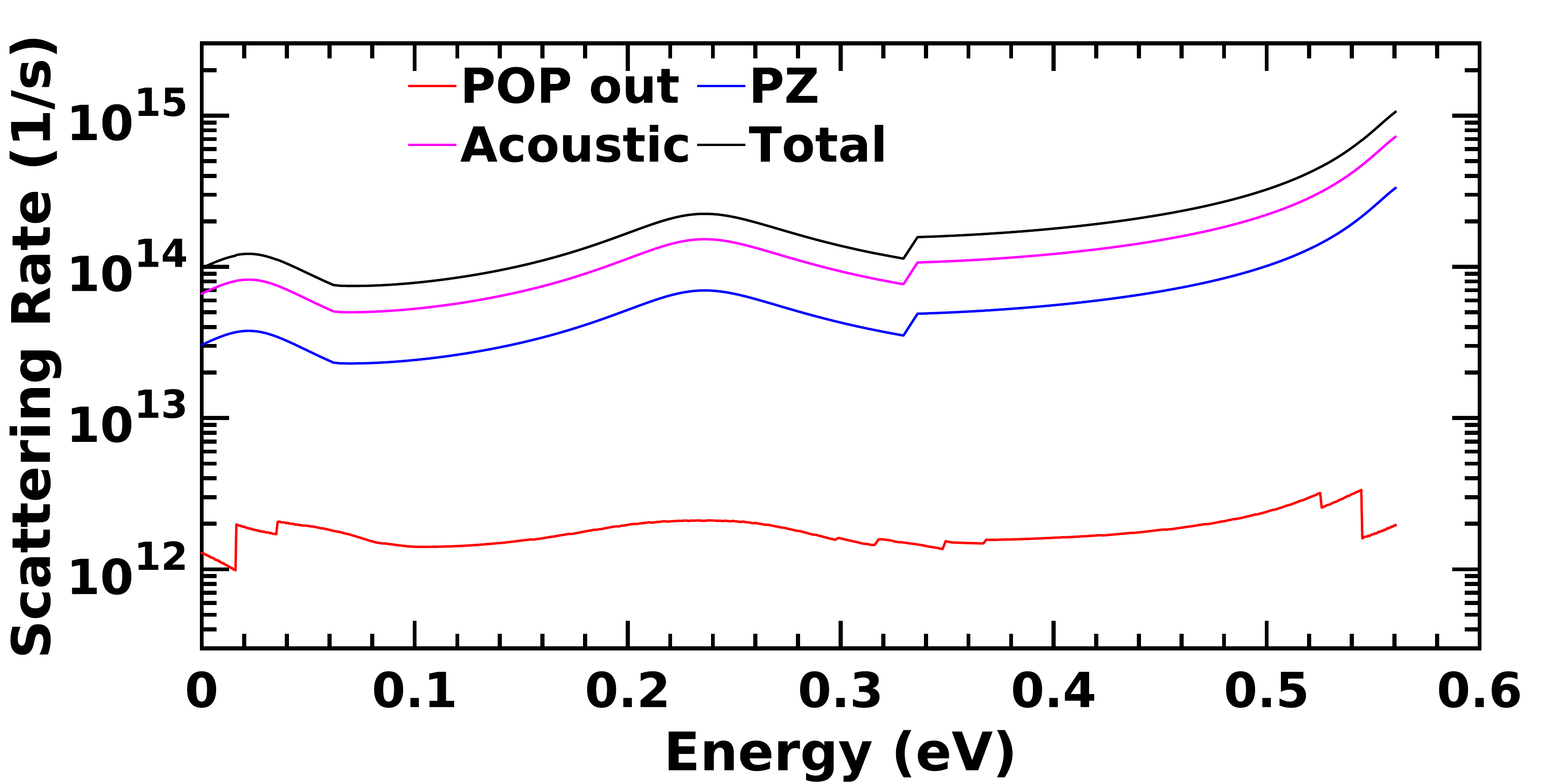}}
    \hfill
    \subfigure[$ $]{\includegraphics[width=80mm,height=60mm]{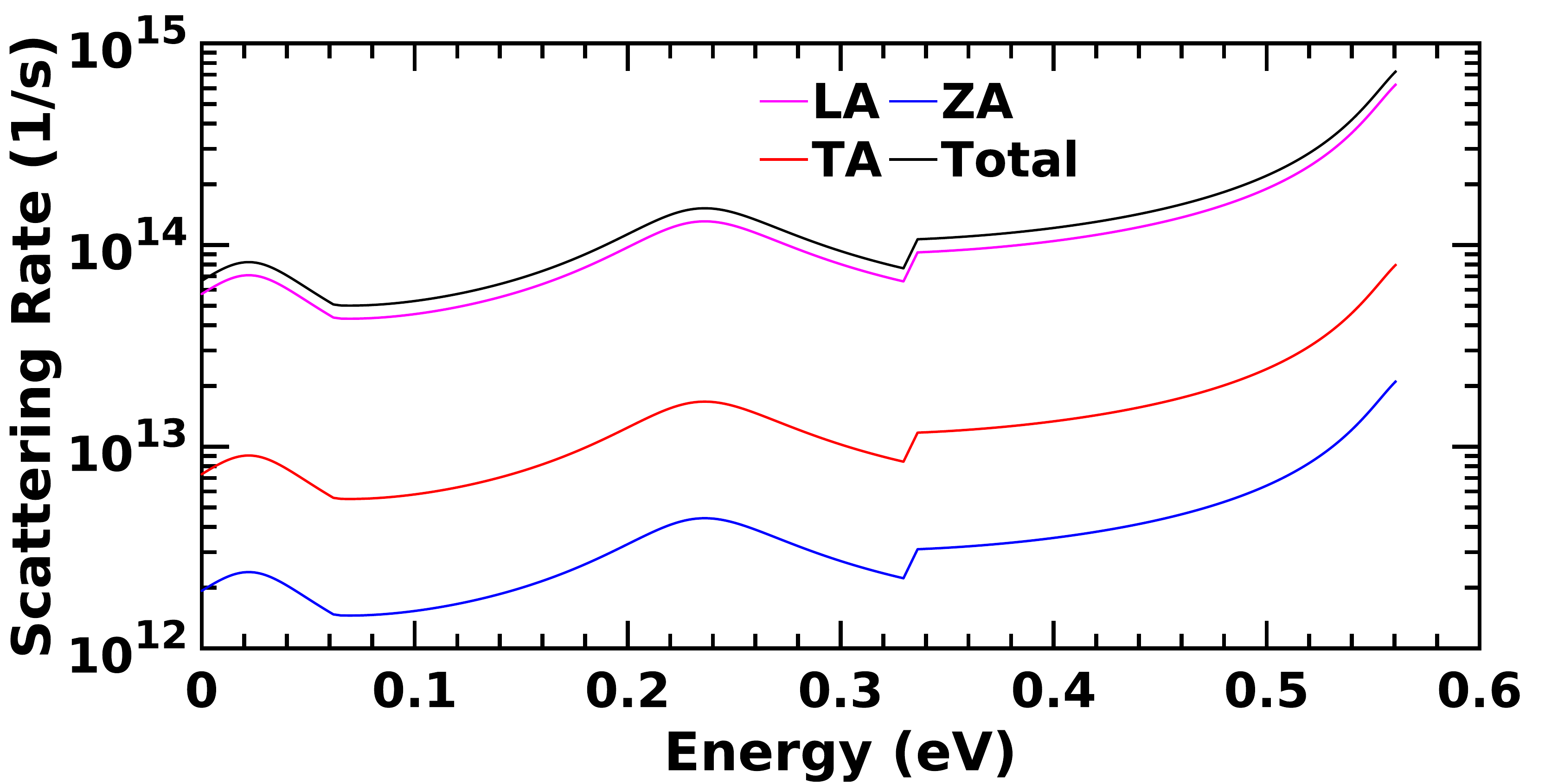}}
    \caption{Scattering Rate vs energy at $300 $ K }
    \label{scattering_rate}
\end{figure} 

\begin{figure}[hbt!]
    \centering
    \subfigure[$ $]{\includegraphics[width=80mm,height=60mm]{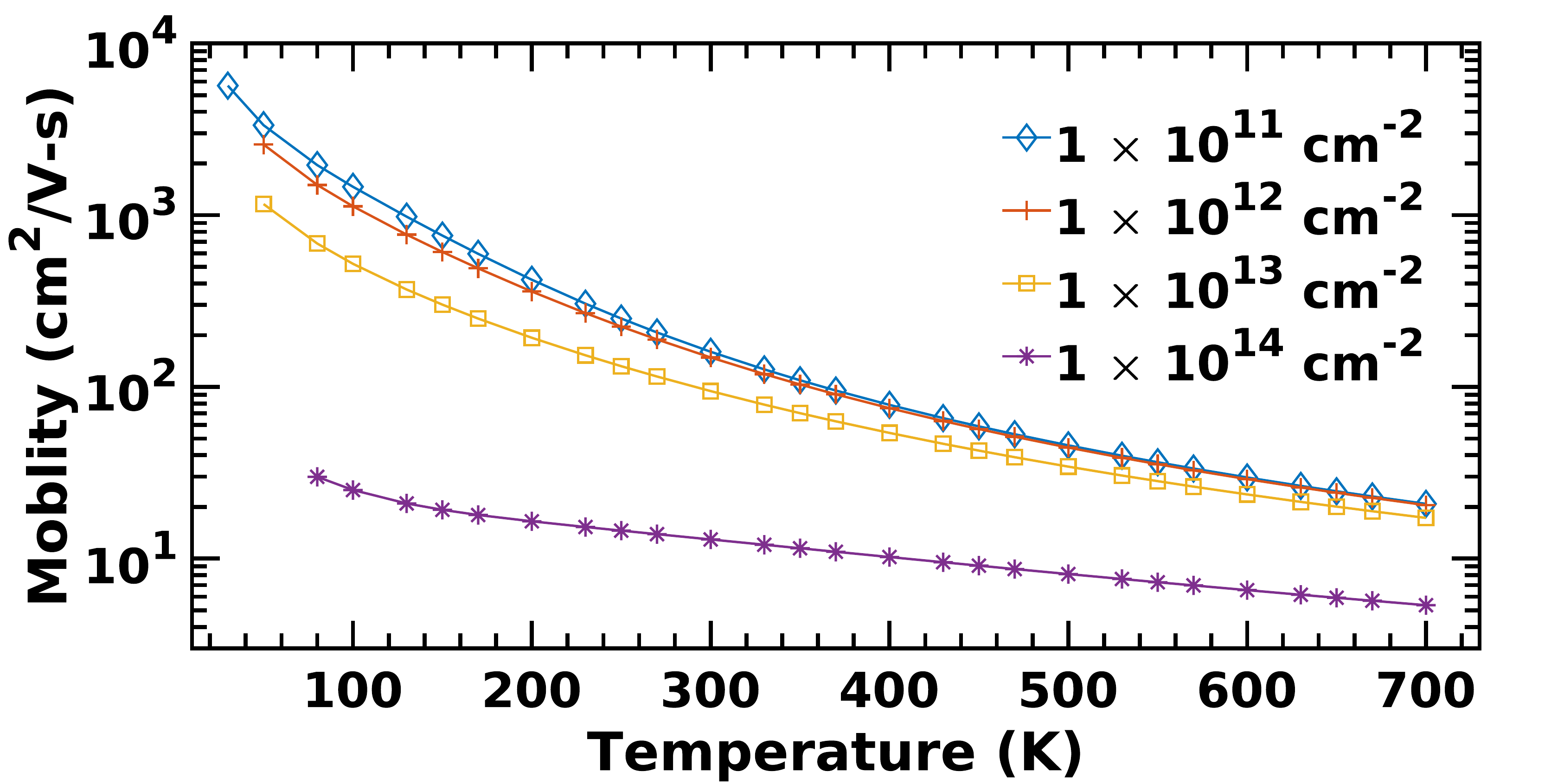}}
    \hfill
    \subfigure[$ $]{\includegraphics[width=80mm,height=60mm]{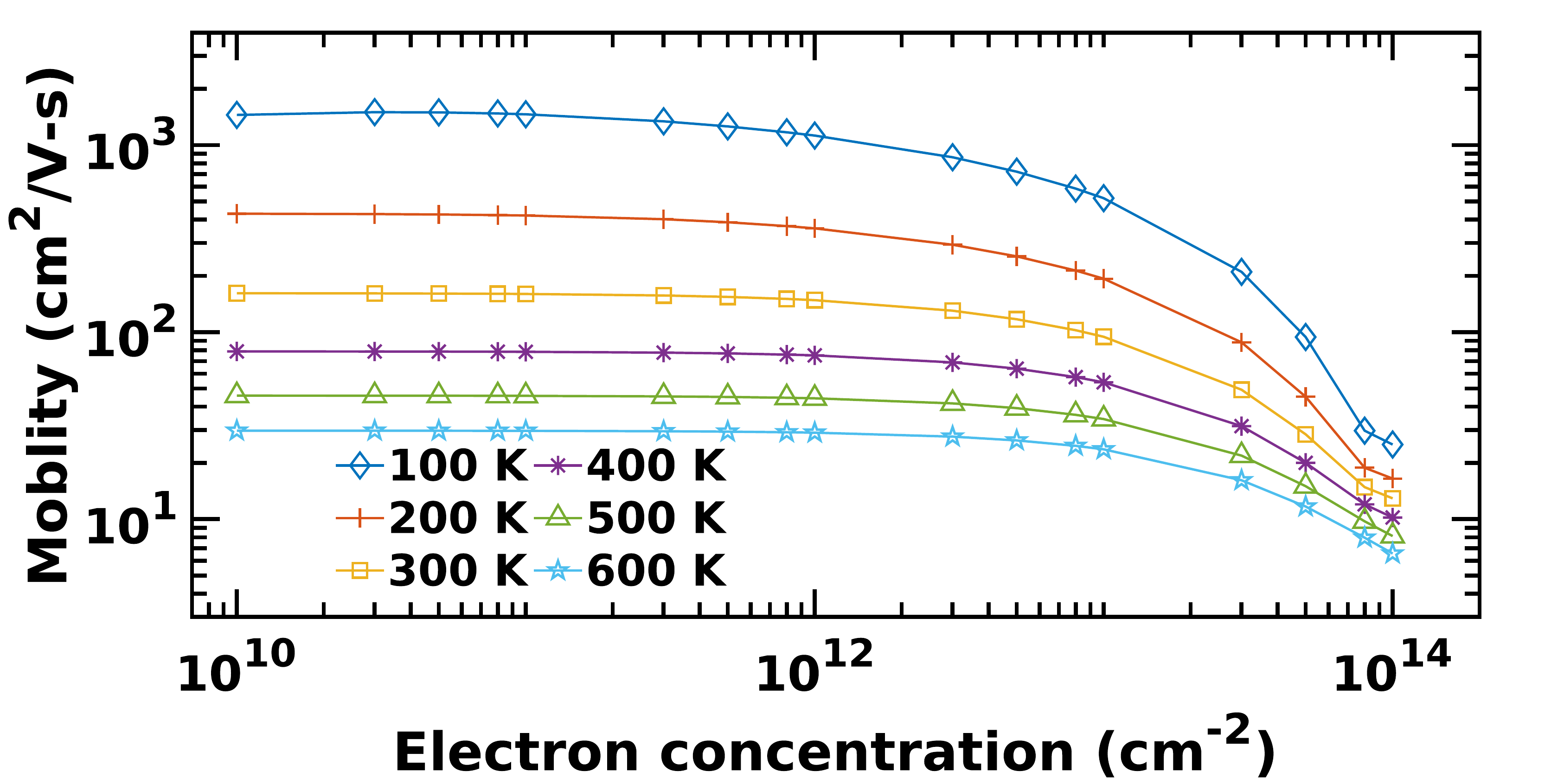}}
    \caption{(a)Variation of mobility with temperature (b) Variation of mobility with doping concentration.}
    \label{mobility}
\end{figure}

\begin{figure}[hbt!]
    \centering
    \subfigure[]{\includegraphics[width=80mm,height=60mm]{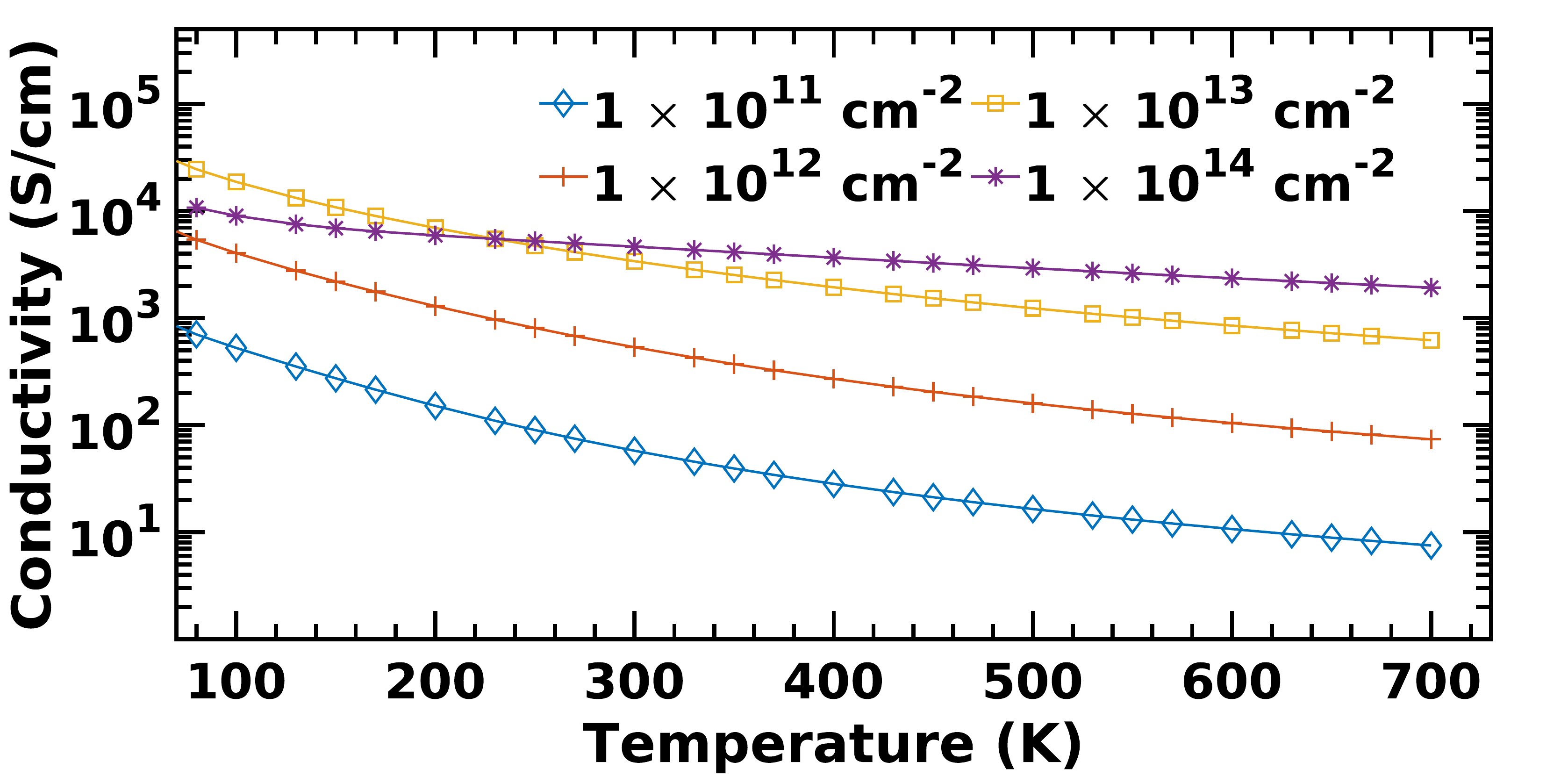}}
    \hfill
    \subfigure[]{\includegraphics[width=80mm,height=60mm]{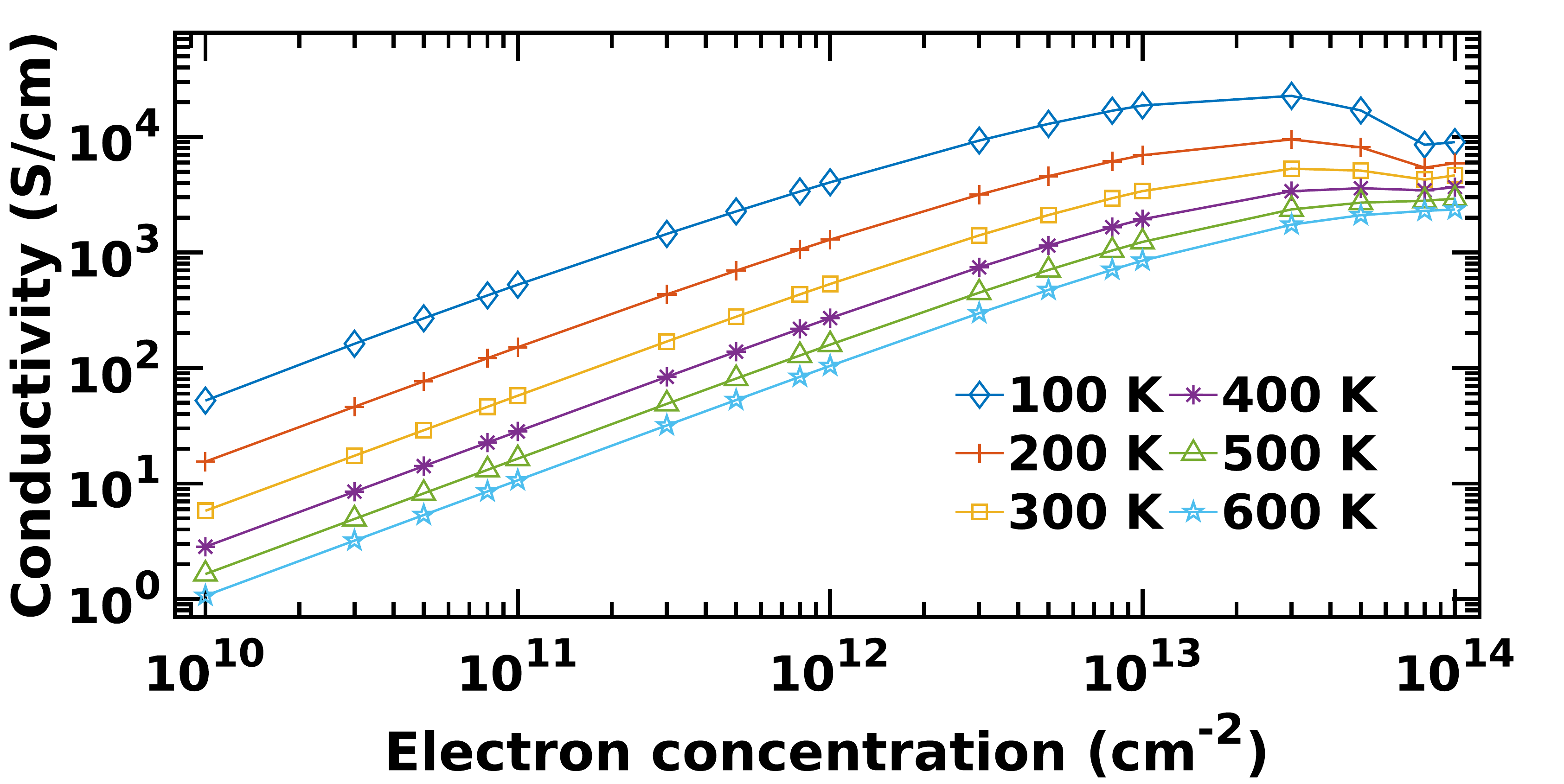}}
    \caption{(a) Variation of conductivity with temperature (b) Variation of  conductivity with doping concentration}
    \label{conductivity}
\end{figure}

\begin{figure}[hbt!]
    \centering
    \subfigure[$ $]{\includegraphics[width=80mm,height=60mm]{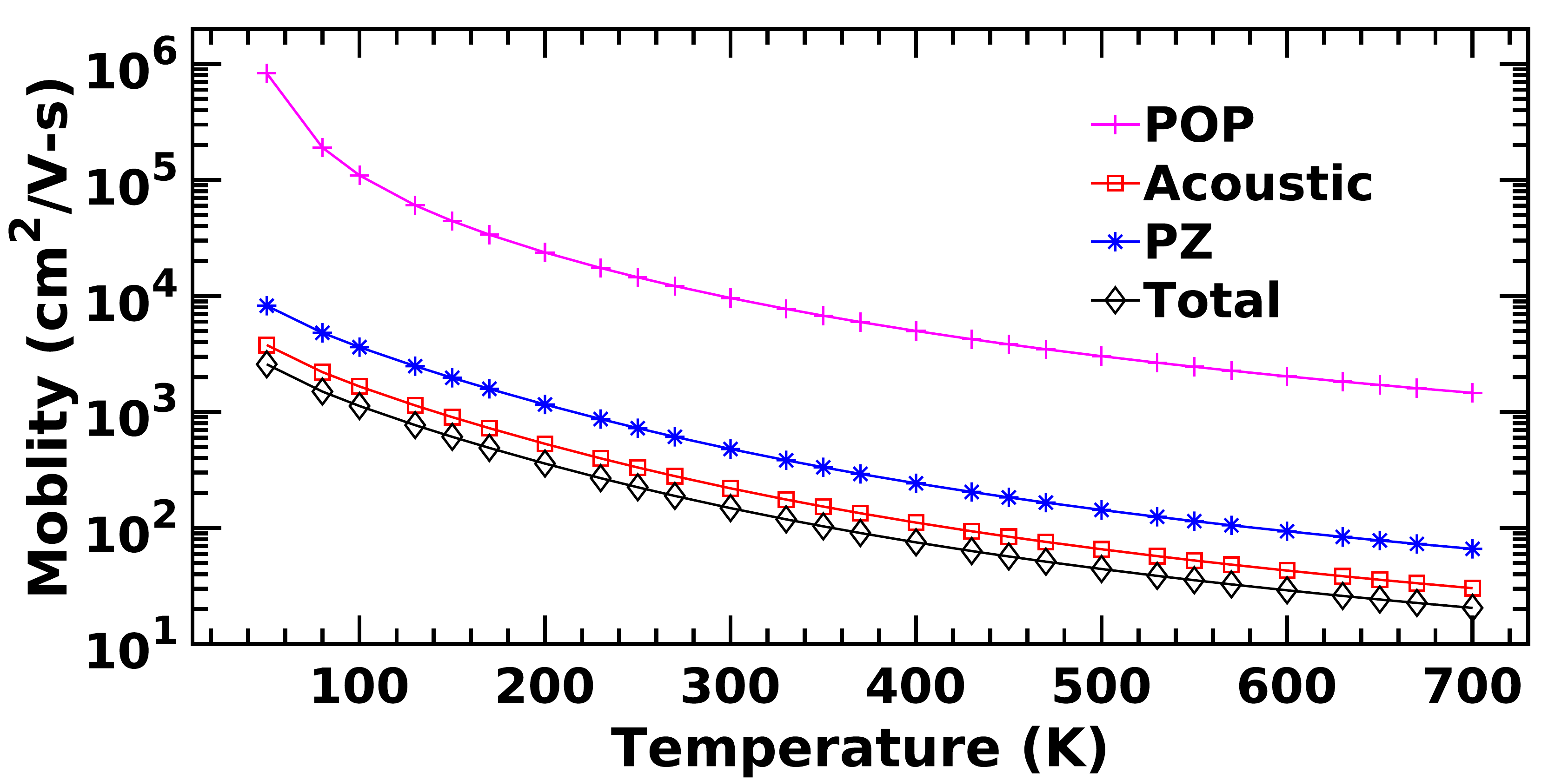}}
    \hfill
    \subfigure[$ $]{\includegraphics[width=80mm,height=60mm]{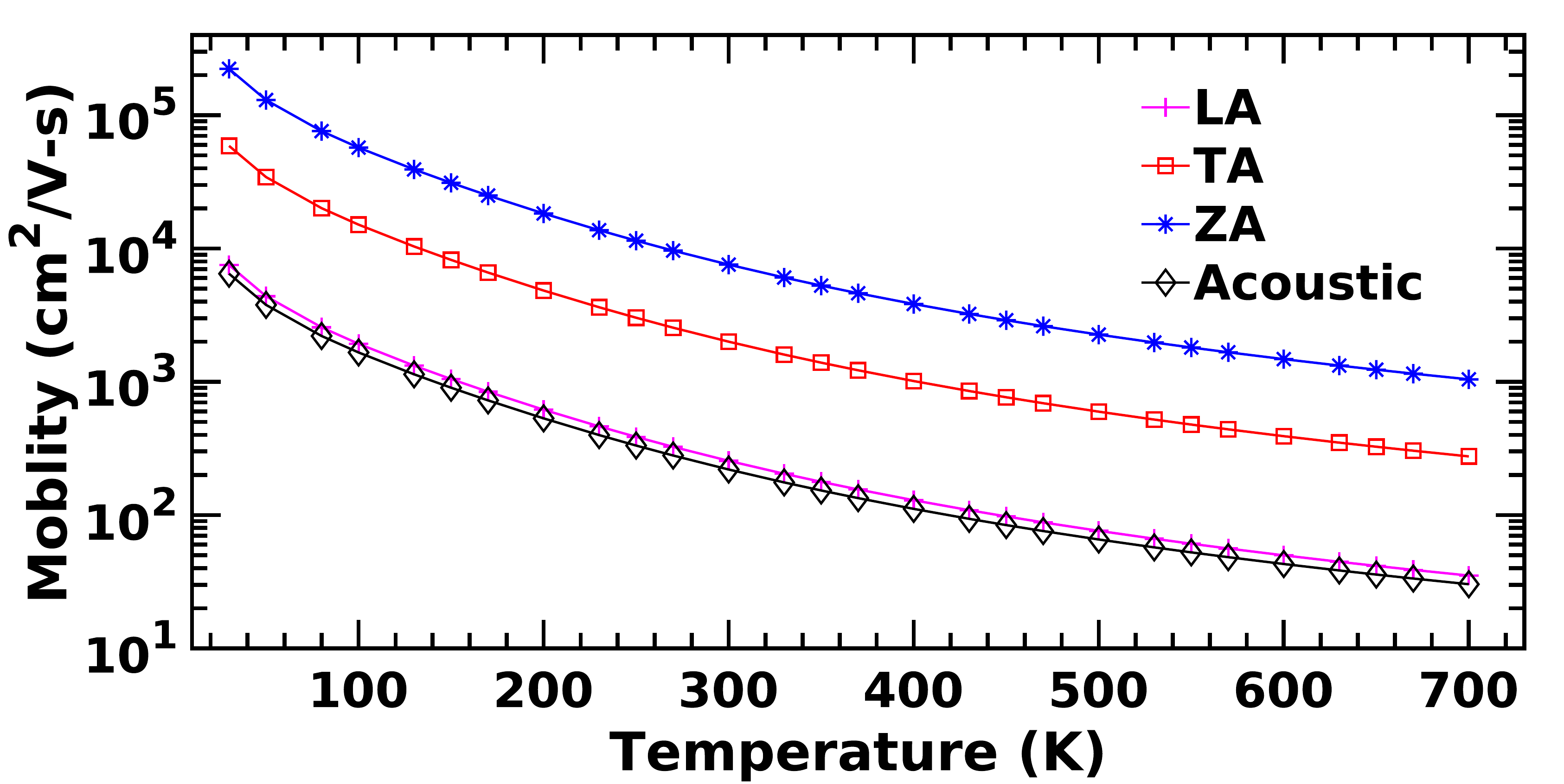}}
    \caption{(a)Contribution of mobility due to polar optical phonons, acoustic phonons and piezoelectric scattering (b) due to LA, TA and ZA acoustic phonon mode with temperature.}
    \label{mo-co}
\end{figure} 
\begin{figure}[hbt!]
    \centering
    \subfigure[]{\includegraphics[width=80mm,height=60mm]{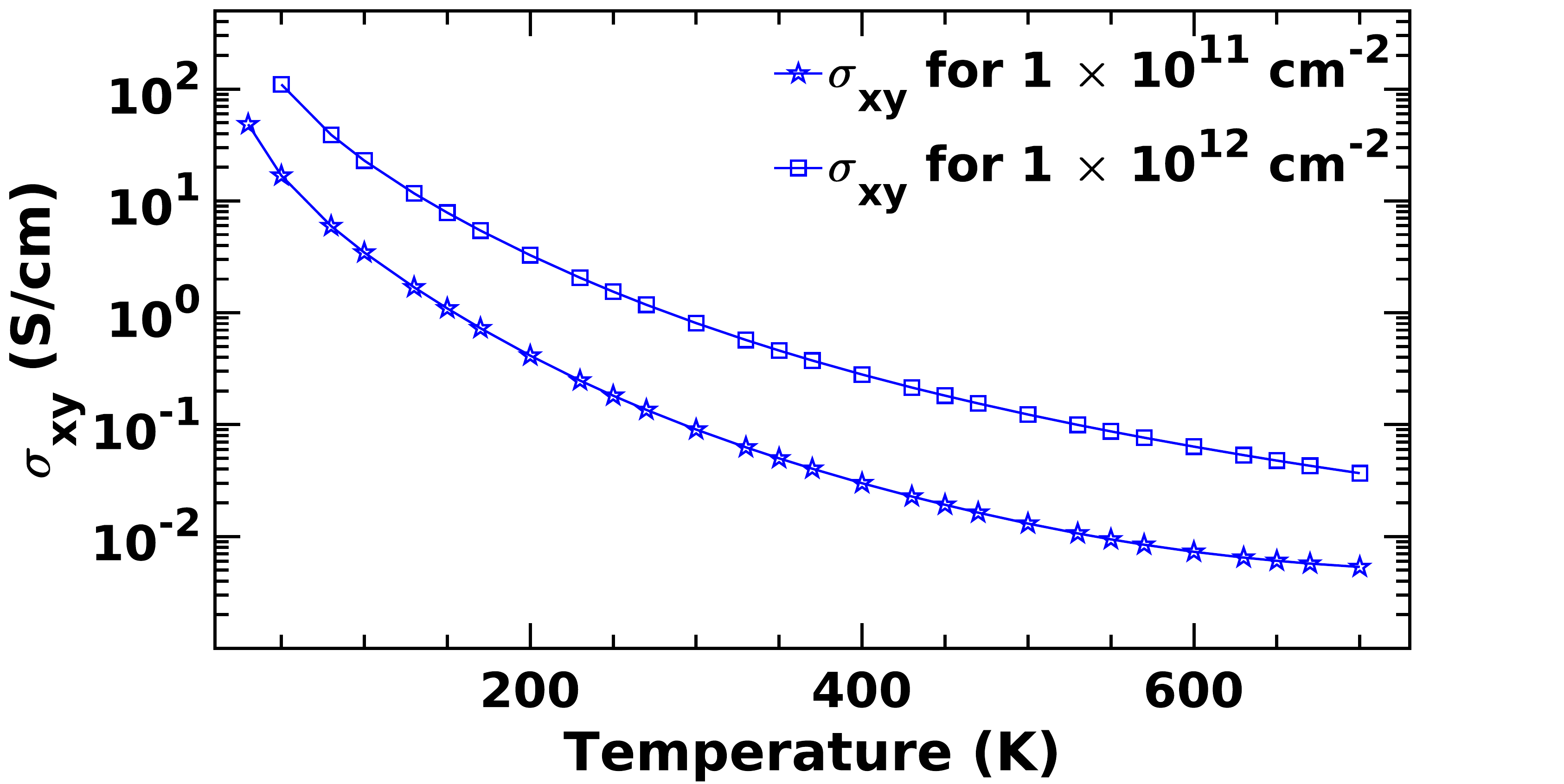}}
    \hfill
    \subfigure[]{\includegraphics[width=80mm,height=60mm]{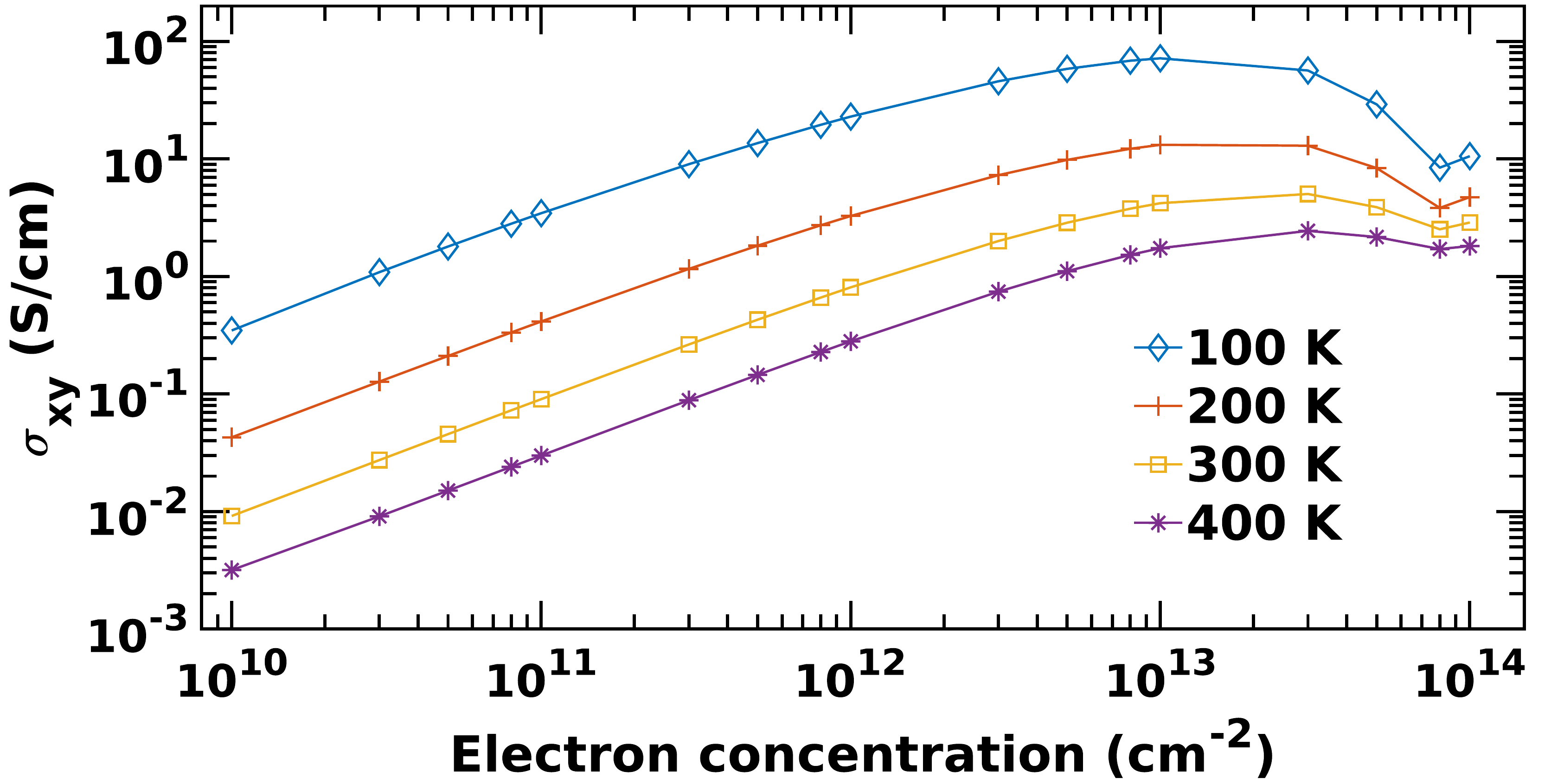}}
    \caption{Variation of Hall conductivity  with (a)temperature and (b) carrier concentration.}
    \label{Hall-cond}
\end{figure}

\begin{figure}[hbt!]
    \centering
    \subfigure[]{\includegraphics[width=80mm,height=60mm]{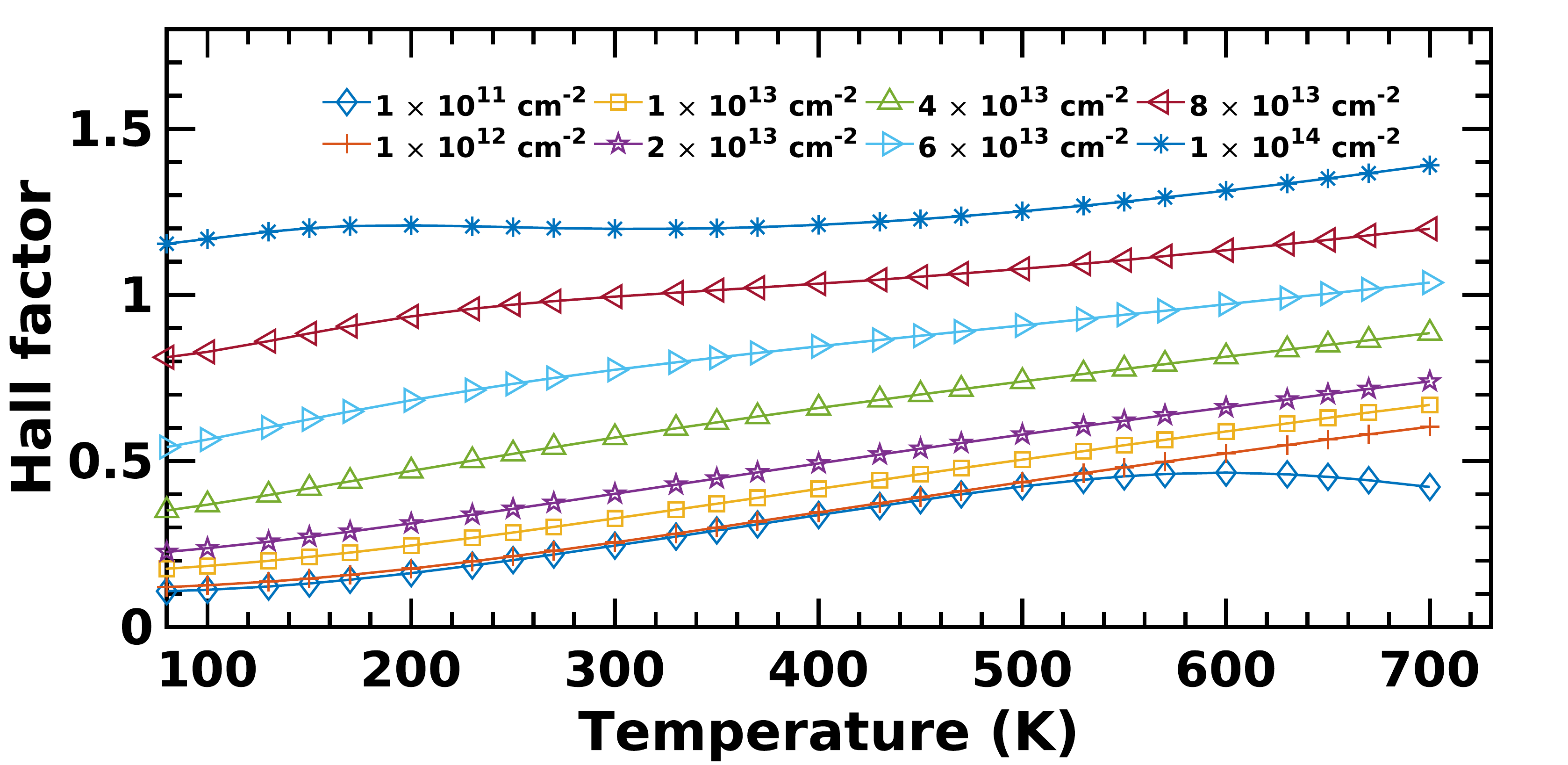}}
    \subfigure[]{\includegraphics[width=80mm,height=60mm]{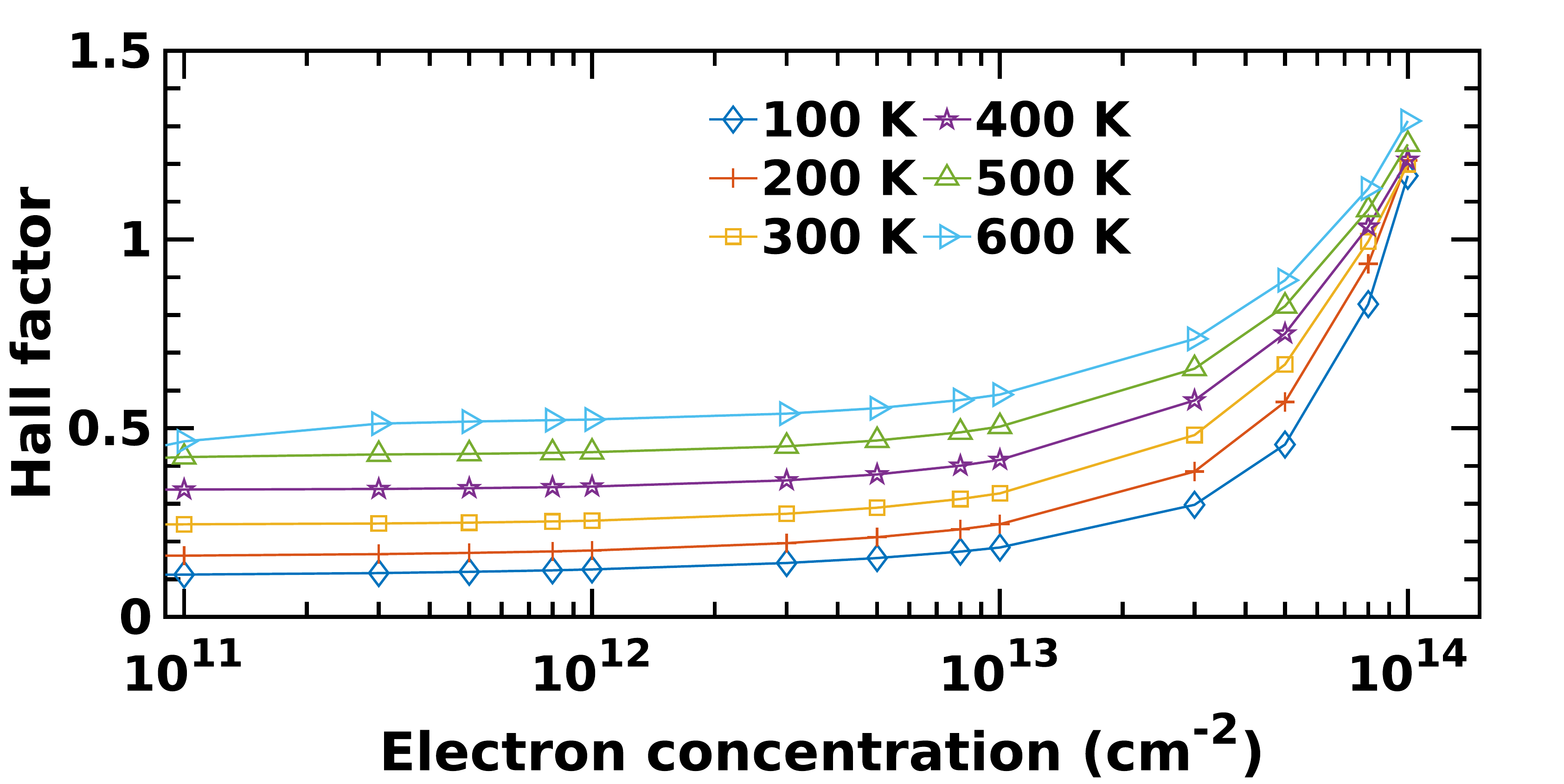}}
    \caption{(a) Variation of Hall factor with temperature (b) Variation of Hall factor with concentration }
    \label{hall_factor}
\end{figure}

\begin{figure}[hbt!]
    \centering
    \subfigure[]{\includegraphics[scale=0.75]{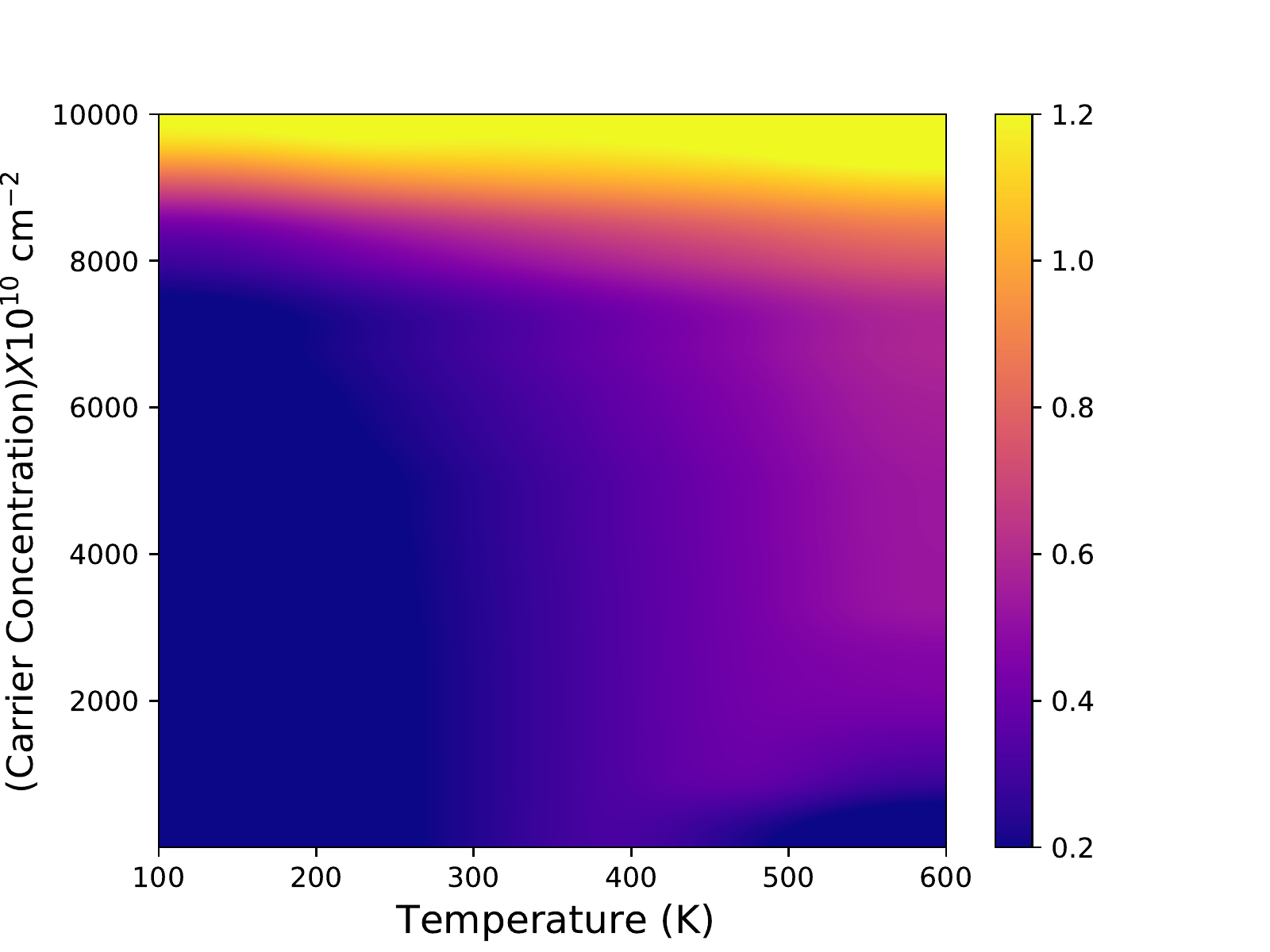}}
    \subfigure[]{\includegraphics[scale=0.75]{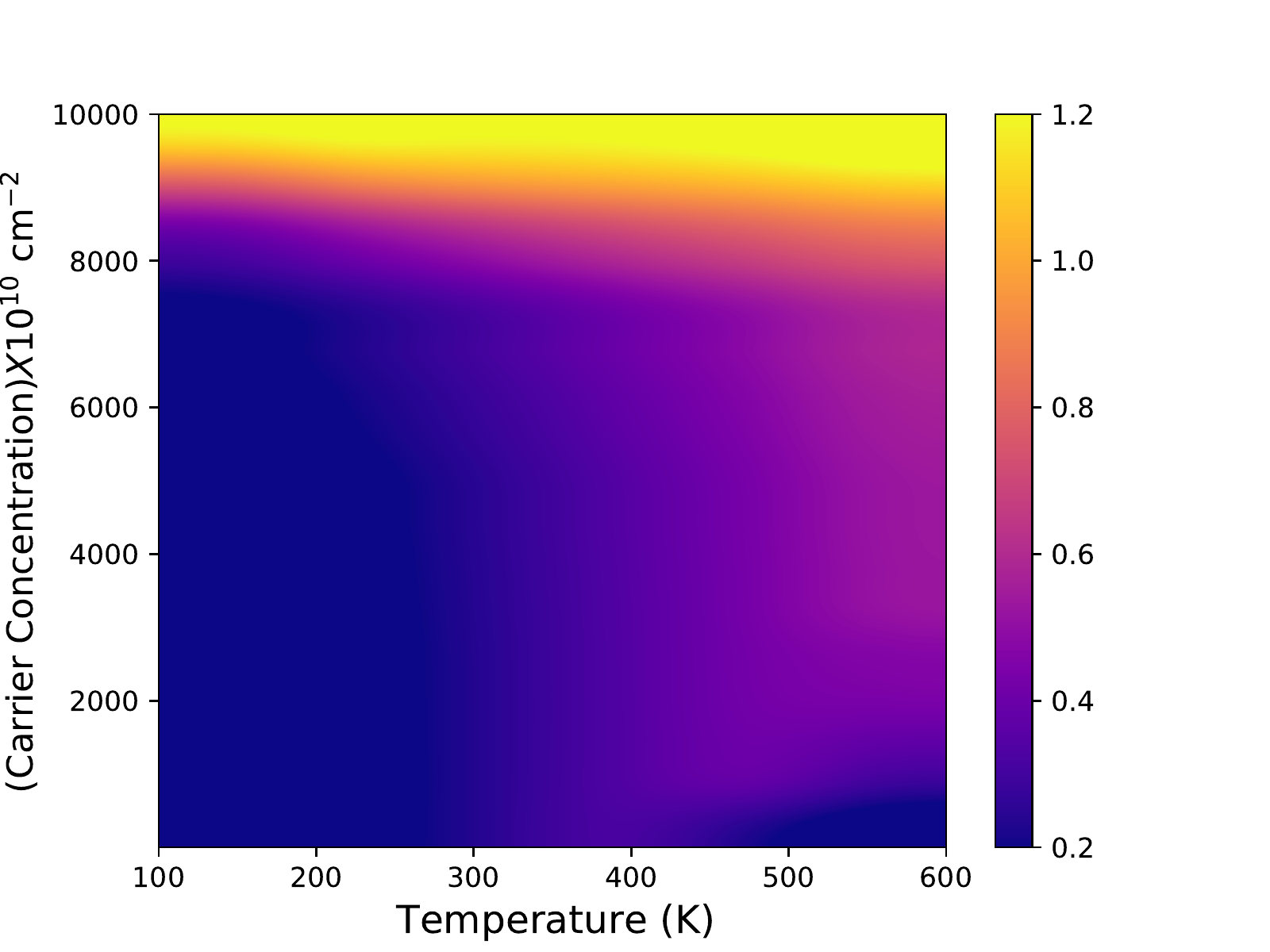}}
    \caption{Hall scattering factor as a function of temperature and carrier concentration at (a)0.4T and (b)0.8T}
    \label{Phase}
\end{figure}
\end{document}